\documentclass[a4paper,12pt]{article}
\usepackage{amsmath,amssymb}
\begin{document}

\begin{center}
{\bf NOTES ON RENORMALIZATION}
\end{center}

\vspace*{-.1cm}

\begin{center}
{\sc A.\,A. Vladimirov} \\[.1cm]
{\em BLTP, JINR, Dubna, Moscow region 141980, Russia \\
alvladim@thsun1.jinr.ru \\
http://thsun1.jinr.ru/{\char126}alvladim/qft.html}
\end{center}

\vspace{.7cm}

{\small
We outline the proofs of several principal statements in conventional
renormalization theory. This may be of some use in the light of new trends
and new techniques (Hopf algebras, etc.) recently introduced in the field.
}

\vspace{1cm}

The following statements form a backbone of renormalization theory
\cite{Co}\,:
\begin{itemize}
  \item[(\textbf R)] In renormalizable models, all ultraviolet divergences
are removed by the use of a special recursive subtraction procedure
($R$-operation)\,.
  \item[(\textbf C)] This operation is completely equivalent to adding
the properly chosen local singular counterterms to the Lagrangian.
  \item[(\textbf G)] A freedom in the renormalization procedure is
precisely a possibility to change parameters in the
Lagrangian, via the renormalization group transformations.
\item[(\textbf S)] This freedom (renormalization-scheme dependence) cancels
out entirely in observable quantities, if expressed in proper variables.
\end{itemize}

Our aim here is to recall generic proofs for these statements. This may
prove to be useful, especially when comparing conventional methods with the
new techniques (Hopf algebras, etc.) recently proposed \cite{CK}.
Addressing the conceptual matter only, we choose to deal with
the simplest cases where, nevertheless, the principal ideas of renormalization
can be demonstrated in unabridged form. For example, one can imagine a
massless theory with a single coupling constant $g$\,, like $g\varphi^3$
in 6 dimensions, additionally `quenched' by discarding all self-energy
(propagator-type) subgraphs. Of course, such a model can be non-unitary as
quantum field theory, but it is fairly well renormalizable, with only
3-point graphs being divergent, and without overlapping divergences.

We begin with the statement ($\mathbf R$)\,. Recursive nature of the
$R$-operation is well understood from the following picture
(empty circle designates the \emph{full} 3-\emph{vertex}, i.e., formal sum
of all graphs having three external legs,
with proper combinatoric factors)\,:
\begin{equation}
\label{R}
R \
\parbox{2.3cm}{
\begin{picture}(60,60)
\put(20,30){\circle{14}}
\put(40,10){\circle{14}}
\put(40,50){\circle{14}}
\put(0,30){\line(1,0){13}}
\put(47,50){\line(1,0){13}}
\put(47,10){\line(1,0){13}}
\put(25,35){\line(1,1){10}}
\put(25,25){\line(1,-1){10}}
\put(40,17){\line(0,1){26.5}}
\end{picture}
} = \ (1-K) \
\parbox{2.3cm}{
\begin{picture}(60,60)
\put(20,30){\circle{14}}
\put(40,10){\circle{14}}
\put(40,50){\circle{14}}
\put(0,30){\line(1,0){13}}
\put(47,50){\line(1,0){13}}
\put(47,10){\line(1,0){13}}
\put(25,35){\line(1,1){10}}
\put(25,25){\line(1,-1){10}}
\put(40,17){\line(0,1){26.5}}
\put(16.7,26){R}
\put(36.7,6){R}
\put(36.7,46){R}
\end{picture}
}\,, \ \ \ \ \ \ \ \ \ \ \
\parbox{1.5cm}{
\begin{picture}(35,60)
\put(20,30){\circle{14}}
\put(0,30){\line(1,0){13}}
\put(25,35){\line(1,1){10}}
\put(25,25){\line(1,-1){10}}
\put(16.7,26){R}
\end{picture}
} \doteq \ R\
\parbox{1.5cm}{
\begin{picture}(35,60)
\put(20,30){\circle{14}}
\put(0,30){\line(1,0){13}}
\put(25,35){\line(1,1){10}}
\put(25,25){\line(1,-1){10}}
\end{picture}
}\,,
\end{equation}
with $K$ being a projector, $K^2=K$\,, picking up the singular
(and $1\!-\!K$ the regular) terms in the corresponding regularization
parameter. In a word, to renormalize a graph one has to renormalize its
subgraphs and then subtract the remaining superficial singularity.
Of course, this procedure terminates for any given graph. But there arises a
nontrivial question: how to make sure that a singular contribution we thus
obtain is local, i.e., does not depend on (or is at most polynomial in)
external momenta? This is crucial for renormalizability, because subtracting
nonlocal quantities cannot be interpreted (and so justified) as adding
counterterms to Lagrangian. As a rule, the proof of locality heavily depends
on the regularization prescription used, and, thus, not so much
can be said in a general language we are using here. However, the following
trick often works: try to find an operation $D$ (say, differentiation
with respect to external momenta \cite{Co}) which eliminates
superficial divergences and commutes with $K$\,. Then $K\!\circ D$
yields zero when applied to a graph which diverges only superficially,
whereas $D\circ K$ differentiates the corresponding counterterm, with zero
result, thus demonstrating its locality.

Let us now proceed with (\textbf{C})\,. The basic reason why recursive
subtractions are equivalent to counterterms lies in the auto-recursive
nature of the perturbation series itself: an insertion of the full 3-vertex
instead of some elementary vertex in a given 3-point graph produces
(a subset of) the graphs of the same full 3-vertex, with correct
combinatoric coefficients. Namely, the following identity holds:
\begin{equation}
\label{autorec}
\parbox{1.5cm}{
\begin{picture}(35,60)
\put(20,30){\circle{14}}
\put(0,30){\line(1,0){13}}
\put(25,35){\line(1,1){10}}
\put(25,25){\line(1,-1){10}}
\end{picture}
} =
\parbox{1.2cm}{
\begin{picture}(30,60)
\put(20,30){\circle*{3}}
\put(5,30){\line(1,0){15}}
\put(20,30){\line(1,1){10}}
\put(20,30){\line(1,-1){10}}
\end{picture}
} + \
\parbox{2.3cm}{
\begin{picture}(60,60)
\put(20,30){\circle{14}}
\put(40,10){\circle{14}}
\put(40,50){\circle{14}}
\put(0,30){\line(1,0){13}}
\put(47,50){\line(1,0){13}}
\put(47,10){\line(1,0){13}}
\put(25,35){\line(1,1){10}}
\put(25,25){\line(1,-1){10}}
\put(40,17){\line(0,1){26.5}}
\end{picture}
} + \ \frac{1}{2} \
\parbox{3.7cm}{
\begin{picture}(100,60)
\put(20,30){\circle{14}}
\put(40,10){\circle{14}}
\put(40,50){\circle{14}}
\put(80,10){\circle{14}}
\put(80,50){\circle{14}}
\put(87,50){\line(1,0){13}}
\put(87,10){\line(1,0){13}}
\put(0,30){\line(1,0){13}}
\put(47,50){\line(1,0){26}}
\put(47,10){\line(1,0){26}}
\put(25,35){\line(1,1){10}}
\put(25,25){\line(1,-1){10}}
\put(45,15){\line(1,1){12}}
\put(75,45){\line(-1,-1){12}}
\put(45,45){\line(1,-1){30}}
\put(63.5,27){\oval(15,15)[tl]}
\end{picture}
} + \ \ldots
\end{equation}
Here each (except first) term of the sum in the r.h.s. is a `primitive'
(only superficially divergent) graph with all its vertices replaced by
full 3-vertices. The sets of graphs present in both sides of (\ref{autorec})
evidently coincide. The combinatoric factors also agree, due to the same
symmetry features of the elementary and the full 3-vertex.

We now \emph{define} a (minimally) renormalized 3-vertex as follows: the
coupling constant $g$\,, at each its occurence in the perturbative expansion
of (unrenormalized) full 3-vertex, should be replaced by a series
$g_B=g+\mathcal{O}(g^3)$ (denoted by a black circle below) with the terms
$\mathcal{O}(g^3)$ assumed to be singular in a regularization parameter:
\begin{equation}
\label{ren}
\parbox{1.2cm}{
\begin{picture}(30,60)
\put(20,30){\circle*{3}}
\put(5,30){\line(1,0){15}}
\put(20,30){\line(1,1){10}}
\put(20,30){\line(1,-1){10}}
\end{picture}
} \rightarrow \
\parbox{1.2cm}{
\begin{picture}(30,60)
\put(20,30){\circle*{10}}
\put(4,30){\line(1,0){15}}
\put(20,30){\line(1,1){10}}
\put(20,30){\line(1,-1){10}}
\end{picture}
} \ \ \ \ \ \ \ \Longrightarrow \ \ \ \ \ \ \
\parbox{1.5cm}{
\begin{picture}(35,60)
\put(20,30){\circle{14}}
\put(0,30){\line(1,0){13}}
\put(25,35){\line(1,1){10}}
\put(25,25){\line(1,-1){10}}
\end{picture}
} \rightarrow \ \,
\parbox{1.5cm}{
\begin{picture}(35,60)
\put(20,30){\circle{14}}
\put(0,30){\line(1,0){13}}
\put(25,35){\line(1,1){10}}
\put(25,25){\line(1,-1){10}}
\put(16.7,26){R}
\end{picture}
}\!\equiv \
R\
\parbox{1.5cm}{
\begin{picture}(35,60)
\put(20,30){\circle{14}}
\put(0,30){\line(1,0){13}}
\put(25,35){\line(1,1){10}}
\put(25,25){\line(1,-1){10}}
\end{picture}
}
\end{equation}
By our assumption,
\begin{equation}
\label{K1}
K
\parbox{1.2cm}{
\begin{picture}(30,60)
\put(20,30){\circle*{10}}
\put(4,30){\line(1,0){15}}
\put(20,30){\line(1,1){10}}
\put(20,30){\line(1,-1){10}}
\end{picture}
} =
\parbox{1.2cm}{
\begin{picture}(30,60)
\put(20,30){\circle*{10}}
\put(4,30){\line(1,0){15}}
\put(20,30){\line(1,1){10}}
\put(20,30){\line(1,-1){10}}
\end{picture}
} - \,
\parbox{1.2cm}{
\begin{picture}(30,60)
\put(20,30){\circle*{3}}
\put(5,30){\line(1,0){15}}
\put(20,30){\line(1,1){10}}
\put(20,30){\line(1,-1){10}}
\end{picture}
}\ , \ \ \ \ \ \ \ \ \ \
K \,
\parbox{1.5cm}{
\begin{picture}(35,60)
\put(20,30){\circle{14}}
\put(0,30){\line(1,0){13}}
\put(25,35){\line(1,1){10}}
\put(25,25){\line(1,-1){10}}
\put(16.7,26){R}
\end{picture}
} = \ 0\ .
\end{equation}
Thus, we lay counterterms in the basis of renormalization.
Now our goal is twofold: to present an explicit formula for counterterms,
and to relate the recipe (\ref{ren}) with $R$-operation.
Both aims are achieved simultaneously: eq.\,(\ref{autorec}),
being identical in $g$\,, remains an identity after
the formal substitution $g\rightarrow g_B$\,, so
\begin{equation}
\label{main}
\parbox{1.5cm}{
\begin{picture}(35,60)
\put(20,30){\circle{14}}
\put(0,30){\line(1,0){13}}
\put(25,35){\line(1,1){10}}
\put(25,25){\line(1,-1){10}}
\put(16.7,26){R}
\end{picture}
} =
\parbox{1.2cm}{
\begin{picture}(30,60)
\put(20,30){\circle*{10}}
\put(4,30){\line(1,0){15}}
\put(20,30){\line(1,1){10}}
\put(20,30){\line(1,-1){10}}
\end{picture}
} + \
\parbox{2.3cm}{
\begin{picture}(60,60)
\put(20,30){\circle{14}}
\put(40,10){\circle{14}}
\put(40,50){\circle{14}}
\put(0,30){\line(1,0){13}}
\put(47,50){\line(1,0){13}}
\put(47,10){\line(1,0){13}}
\put(25,35){\line(1,1){10}}
\put(25,25){\line(1,-1){10}}
\put(40,17){\line(0,1){26.5}}
\put(16.7,26){R}
\put(36.7,6){R}
\put(36.7,46){R}
\end{picture}
} + \ \frac{1}{2} \
\parbox{3.7cm}{
\begin{picture}(100,60)
\put(20,30){\circle{14}}
\put(40,10){\circle{14}}
\put(40,50){\circle{14}}
\put(80,10){\circle{14}}
\put(80,50){\circle{14}}
\put(87,50){\line(1,0){13}}
\put(87,10){\line(1,0){13}}
\put(0,30){\line(1,0){13}}
\put(47,50){\line(1,0){26}}
\put(47,10){\line(1,0){26}}
\put(25,35){\line(1,1){10}}
\put(25,25){\line(1,-1){10}}
\put(45,15){\line(1,1){12}}
\put(75,45){\line(-1,-1){12}}
\put(45,45){\line(1,-1){30}}
\put(63.5,27){\oval(15,15)[tl]}
\put(16.7,26){R}
\put(36.7,6){R}
\put(36.7,46){R}
\put(76.7,6){R}
\put(76.7,46){R}
\end{picture}
} + \ \ldots
\end{equation}
Using (\ref{K1}) immediately yields
\begin{multline}
\label{counter}
\parbox{1.2cm}{
\begin{picture}(30,60)
\put(20,30){\circle*{10}}
\put(4,30){\line(1,0){15}}
\put(20,30){\line(1,1){10}}
\put(20,30){\line(1,-1){10}}
\end{picture}
} =
\parbox{1.2cm}{
\begin{picture}(30,60)
\put(20,30){\circle*{3}}
\put(5,30){\line(1,0){15}}
\put(20,30){\line(1,1){10}}
\put(20,30){\line(1,-1){10}}
\end{picture}
} - K\left(
\parbox{2.3cm}{
\begin{picture}(60,60)
\put(20,30){\circle{14}}
\put(40,10){\circle{14}}
\put(40,50){\circle{14}}
\put(0,30){\line(1,0){13}}
\put(47,50){\line(1,0){13}}
\put(47,10){\line(1,0){13}}
\put(25,35){\line(1,1){10}}
\put(25,25){\line(1,-1){10}}
\put(40,17){\line(0,1){26.5}}
\put(16.7,26){R}
\put(36.7,6){R}
\put(36.7,46){R}
\end{picture}
} + \ \frac{1}{2} \
\parbox{3.7cm}{
\begin{picture}(100,60)
\put(20,30){\circle{14}}
\put(40,10){\circle{14}}
\put(40,50){\circle{14}}
\put(80,10){\circle{14}}
\put(80,50){\circle{14}}
\put(87,50){\line(1,0){13}}
\put(87,10){\line(1,0){13}}
\put(0,30){\line(1,0){13}}
\put(47,50){\line(1,0){26}}
\put(47,10){\line(1,0){26}}
\put(25,35){\line(1,1){10}}
\put(25,25){\line(1,-1){10}}
\put(45,15){\line(1,1){12}}
\put(75,45){\line(-1,-1){12}}
\put(45,45){\line(1,-1){30}}
\put(63.5,27){\oval(15,15)[tl]}
\put(16.7,26){R}
\put(36.7,6){R}
\put(36.7,46){R}
\put(76.7,6){R}
\put(76.7,46){R}
\end{picture}
} + \ \ldots \right) \\
\doteq \
\parbox{1.2cm}{
\begin{picture}(30,60)
\put(20,30){\circle*{3}}
\put(5,30){\line(1,0){15}}
\put(20,30){\line(1,1){10}}
\put(20,30){\line(1,-1){10}}
\end{picture}
} + S_K\left(
\parbox{2.3cm}{
\begin{picture}(60,60)
\put(20,30){\circle{14}}
\put(40,10){\circle{14}}
\put(40,50){\circle{14}}
\put(0,30){\line(1,0){13}}
\put(47,50){\line(1,0){13}}
\put(47,10){\line(1,0){13}}
\put(25,35){\line(1,1){10}}
\put(25,25){\line(1,-1){10}}
\put(40,17){\line(0,1){26.5}}
\end{picture}
} + \ \frac{1}{2} \
\parbox{3.7cm}{
\begin{picture}(100,60)
\put(20,30){\circle{14}}
\put(40,10){\circle{14}}
\put(40,50){\circle{14}}
\put(80,10){\circle{14}}
\put(80,50){\circle{14}}
\put(87,50){\line(1,0){13}}
\put(87,10){\line(1,0){13}}
\put(0,30){\line(1,0){13}}
\put(47,50){\line(1,0){26}}
\put(47,10){\line(1,0){26}}
\put(25,35){\line(1,1){10}}
\put(25,25){\line(1,-1){10}}
\put(45,15){\line(1,1){12}}
\put(75,45){\line(-1,-1){12}}
\put(45,45){\line(1,-1){30}}
\put(63.5,27){\oval(15,15)[tl]}
\end{picture}
} + \ \ldots \right)\,,
\end{multline}
where the notation $S_K$ is used for counterterms.
At last, from (\ref{counter}) and (\ref{main}) we conclude:
\begin{multline}
\label{reneq}
\parbox{1.5cm}{
\begin{picture}(35,60)
\put(20,30){\circle{14}}
\put(0,30){\line(1,0){13}}
\put(25,35){\line(1,1){10}}
\put(25,25){\line(1,-1){10}}
\put(16.7,26){R}
\end{picture}
} =
\parbox{1.2cm}{
\begin{picture}(30,60)
\put(20,30){\circle*{3}}
\put(5,30){\line(1,0){15}}
\put(20,30){\line(1,1){10}}
\put(20,30){\line(1,-1){10}}
\end{picture}
} + \,(1-K)\left(
\parbox{2.3cm}{
\begin{picture}(60,60)
\put(20,30){\circle{14}}
\put(40,10){\circle{14}}
\put(40,50){\circle{14}}
\put(0,30){\line(1,0){13}}
\put(47,50){\line(1,0){13}}
\put(47,10){\line(1,0){13}}
\put(25,35){\line(1,1){10}}
\put(25,25){\line(1,-1){10}}
\put(40,17){\line(0,1){26.5}}
\put(16.7,26){R}
\put(36.7,6){R}
\put(36.7,46){R}
\end{picture}
} + \ \frac{1}{2} \
\parbox{3.7cm}{
\begin{picture}(100,60)
\put(20,30){\circle{14}}
\put(40,10){\circle{14}}
\put(40,50){\circle{14}}
\put(80,10){\circle{14}}
\put(80,50){\circle{14}}
\put(87,50){\line(1,0){13}}
\put(87,10){\line(1,0){13}}
\put(0,30){\line(1,0){13}}
\put(47,50){\line(1,0){26}}
\put(47,10){\line(1,0){26}}
\put(25,35){\line(1,1){10}}
\put(25,25){\line(1,-1){10}}
\put(45,15){\line(1,1){12}}
\put(75,45){\line(-1,-1){12}}
\put(45,45){\line(1,-1){30}}
\put(63.5,27){\oval(15,15)[tl]}
\put(16.7,26){R}
\put(36.7,6){R}
\put(36.7,46){R}
\put(76.7,6){R}
\put(76.7,46){R}
\end{picture}
} + \ \ldots \right) \\
\equiv \ R \
\parbox{1.5cm}{
\begin{picture}(35,60)
\put(20,30){\circle{14}}
\put(0,30){\line(1,0){13}}
\put(25,35){\line(1,1){10}}
\put(25,25){\line(1,-1){10}}
\end{picture}
} =
\parbox{1.2cm}{
\begin{picture}(30,60)
\put(20,30){\circle*{3}}
\put(5,30){\line(1,0){15}}
\put(20,30){\line(1,1){10}}
\put(20,30){\line(1,-1){10}}
\end{picture}
} + R\left(
\parbox{2.3cm}{
\begin{picture}(60,60)
\put(20,30){\circle{14}}
\put(40,10){\circle{14}}
\put(40,50){\circle{14}}
\put(0,30){\line(1,0){13}}
\put(47,50){\line(1,0){13}}
\put(47,10){\line(1,0){13}}
\put(25,35){\line(1,1){10}}
\put(25,25){\line(1,-1){10}}
\put(40,17){\line(0,1){26.5}}
\end{picture}
} + \ \frac{1}{2} \
\parbox{3.7cm}{
\begin{picture}(100,60)
\put(20,30){\circle{14}}
\put(40,10){\circle{14}}
\put(40,50){\circle{14}}
\put(80,10){\circle{14}}
\put(80,50){\circle{14}}
\put(87,50){\line(1,0){13}}
\put(87,10){\line(1,0){13}}
\put(0,30){\line(1,0){13}}
\put(47,50){\line(1,0){26}}
\put(47,10){\line(1,0){26}}
\put(25,35){\line(1,1){10}}
\put(25,25){\line(1,-1){10}}
\put(45,15){\line(1,1){12}}
\put(75,45){\line(-1,-1){12}}
\put(45,45){\line(1,-1){30}}
\put(63.5,27){\oval(15,15)[tl]}
\end{picture}
} + \ \ldots \right)\,,
\end{multline}
\begin{equation}
\label{SK}
S_K\left(
\parbox{2.3cm}{
\begin{picture}(60,60)
\put(20,30){\circle{14}}
\put(40,10){\circle{14}}
\put(40,50){\circle{14}}
\put(0,30){\line(1,0){13}}
\put(47,50){\line(1,0){13}}
\put(47,10){\line(1,0){13}}
\put(25,35){\line(1,1){10}}
\put(25,25){\line(1,-1){10}}
\put(40,17){\line(0,1){26.5}}
\end{picture}
} + \ \ldots \right) = - K\left(
\parbox{2.3cm}{
\begin{picture}(60,60)
\put(20,30){\circle{14}}
\put(40,10){\circle{14}}
\put(40,50){\circle{14}}
\put(0,30){\line(1,0){13}}
\put(47,50){\line(1,0){13}}
\put(47,10){\line(1,0){13}}
\put(25,35){\line(1,1){10}}
\put(25,25){\line(1,-1){10}}
\put(40,17){\line(0,1){26.5}}
\put(16.7,26){R}
\put(36.7,6){R}
\put(36.7,46){R}
\end{picture}
} + \ \ldots \right)\ .
\end{equation}
Now (\ref{counter}), (\ref{SK}) give a recipe for evaluating counterterms
and (\ref{reneq}) for renormalizing arbitrary graphs
(or, rather, the interplay of these relations provides an iterative
procedure for finding both counterterms and renormalized values).
Of course, these are readily recognized as the well-known recursive
prescriptions of $R$-operation in its conventional form.

Now we are in a position to justify the statement (\textbf{G})\,. Since (the
properly organized) subtractions are done by counterterms, any possible
finite (re)subtractions can also be done by finite (non-singular)
counterterms. Thus, for observable quantities, all the difference between
two renormalization schemes boils down to a (finite) redefinition of $g$\,,
\begin{equation}
\label{sch}
\tilde f(\frac{p}{\mu}\,,\,\tilde g\,(g)) = f(\frac{p}{\mu}\,,\,g)\,,
\ \ \ \ \ \ \ \tilde g\,(g) = g + \mathcal{O}(g^3)\,,
\end{equation}
where $p$ represents momenta, and $\mu$ is a \emph{renormalization parameter}
inevitably present in any renormalization procedure
(and denoted here by the same letter in both schemes). Another
form of this relation addresses the case of different $\mu$'s within the
same scheme (for different $\mu$'s imply different renormalization)\,:
\begin{equation}
\label{reninv}
f(\frac{p}{\mu}\,,\,g) = f(\frac{p}{\mu_0}\,,\,g_0)\,, \ \ \ \ \ \ \ \ \ \
g = \bar{g}\,(\frac{\mu}{\mu_0}\,,\,g_0)\,.
\end{equation}
This property is called \emph{renormalization invariance}\,. The
corresponding transformations $\mu_0\rightarrow\mu\,,\,g_0\rightarrow g$\,,
which leave physics intact, form the \emph{renormalization group}\,.
A function $\bar{g}$ is \emph{effective} (or \emph{running}) coupling
constant. It shows how $g$ must change to compensate a change in $\mu$\,,
and enters into the main formula of the renormalization group approach,
\begin{equation}
\label{RG}
f(\frac{p}{\mu}\,,\,g) = f(1\,,\,\bar{g}\,(\frac{p}{\mu}\,,\,g))\,,
\end{equation}
which trades the $p$-dependence of the function $f$ for its dependence
on the (now effectively $p$-dependent) coupling $\bar{g}$\,. Furthermore,
being actually a record of the group transformation law, $\bar{g}$
is subject to standard conditions
\begin{equation}
\label{group}
\bar{g}\,(xy\,,\,g) =  \bar{g}\,(x\,,\, \bar{g}\,(y\,,\,g))\,, \ \ \ \ \ \ \
\bar{g}\,(1\,,\,g) = g\,,
\end{equation}
which are usually cast into differential equations governed by
$\beta$-function:
\begin{equation}
\label{RGdiff}
(x\partial_x - \beta(g)\,\partial_g)\bar{g}\,(x\,,\,g) = 0\,, \ \ \ \ \
x\partial_x\,\bar{g}\,(x\,,\,g) = \beta(\bar{g}\,(x\,,\,g))\,, \ \ \ \ \
\beta(g)\doteq\partial_x\bar{g}\,(x\,,\,g)_{x=1}
\end{equation}

The very existence of a (non-singular) dependence $g\,(\mu)$ given by
(\ref{reninv}) which leaves observable quantities invariant, has interesting
consequences \cite{tH}. Since both $g$ and $\mu$ come into play through
$g_B$ only, one can find the derivative (at $g_B=\text{const}$) \,
$dg/d\log\mu=\beta(g)$ directly from the (singular) expansion of $g_B$
in powers of a regularization parameter $M$\,:
\begin{equation}
\label{gB}
g_B\,(\frac{M}{\mu}\,,\,g) = g + a(g)\log\frac{M}{\mu}
 + b(g)\log^2\frac{M}{\mu} + \ldots\ ,
\end{equation}
\begin{equation}
\notag
\frac{dg}{d\log\mu}
 = -\frac{\partial g_B/\partial\log\mu}{\partial g_B/\partial g}
 = \frac{a(g) + 2b(g)\log\frac{M}{\mu} + \ldots}
  {1 + a'(g)\log\frac{M}{\mu} + \ldots}
 = a(g) + (2b(g) - a'(g))\log\frac{M}{\mu} + \ldots
\end{equation}
The first term in the r.h.s. gives $\beta(g)=a(g)$\,, while other terms are
singular and must cancel out. Thus we obtain $2b(g)=a'(g)$ and other
formulas expressing the coefficient functions at higher powers of
$\log(M/\mu)$ in (\ref{gB}) in terms of $a(g)$\,.

Finally, consider the statement (\textbf{S})\,. From (\ref{sch})
and (\ref{RG}) one can see that the renormalization scheme dependence
of $f$ and $\bar g$\, is governed by the same function $\tilde g\,(g)$.
To show that the two dependences actually cancel, we define
(up to a constant) a function $\psi(g)$\,,
\begin{equation}
\label{psi}
\psi'(g) \doteq \frac{1}{\beta(g)} \ \ \ \ \Longrightarrow \ \ \ \ \
d\log\mu = d\psi(g) \ \ \ \ \Longrightarrow \ \ \ \ \
\log\mu = \psi(g) + \text{const}
\end{equation}
that allows us to rewrite the original definition of $\bar g$ in the form
\begin{equation}
\label{}
\psi\,(\bar{g}\,(\frac{\mu}{\mu_0}\,,\,g_0)) - \psi(g_0)
 = \log\frac{\mu}{\mu_0}
\end{equation}
and introduce (again up to a constant) a new expansion parameter\,
$\log\Lambda$\,:
\begin{equation}
\label{Lambda}
\bar{g}\,(\frac{p}{\mu}\,,\,g) = \psi^{-1}(\log\frac{p}{\mu} + \psi(g))
 = \psi^{-1}(\log\frac{p}{\Lambda})\,, \ \ \ \ \ \ \ \ \
 \Lambda \doteq \mu e^{-\psi(g)}\ .
\end{equation}
The scheme dependence of $\psi$\,, according to the rule adopted in
(\ref{sch})\,: $\mu\rightarrow\mu, \ g\rightarrow\tilde g\,(g)$\,,
is extracted from the last equality in (\ref{psi})\,:
\begin{equation}
\label{psi1}
\tilde\psi(\tilde g\,(g)) = \psi(g) + \text{const}\,, \ \ \ \ \ \ \ \ \
\tilde\psi^{-1}\,(x + \text{const}) = \tilde g\,(\psi^{-1}(x))\,.
\end{equation}
Now we come to a scheme-independent expansion recipe \cite{Vl}:
\begin{gather}
\label{indep}
f(\frac{p}{\mu}\,,\,g) = f(1\,,\,\bar{g}\,(\frac{p}{\mu}\,,\,g))
 = f(1\,,\,\psi^{-1}(\log\frac{p}{\Lambda}))
 = \Phi\,(\log\frac{p}{\Lambda})\,, \\
\label{Phi}
\Phi\,(x) \doteq f(1\,,\,\psi^{-1}(x))\,, \ \ \ \ \ \ \ \ \ \ \
\tilde\Phi\,(x + \text{const}) = \Phi\,(x)\,.
\end{gather}

Let us summarize the whole procedure. One obtains $f$ and $\beta(g)$ by
evaluating Feynman graphs;  $\psi(g)$ is found from (\ref{psi})\,.
The scheme dependences of $f$ (\ref{sch}) and $\psi$
(\ref{psi1}) cancel each other in the function $\Phi$\,,
which is expanded in $\log(p/\Lambda)$ like this:
\begin{equation}
\label{expans}
\Phi\,(\log\frac{p}{\Lambda}) = 1 + \frac{a_1}{\log\frac{p}{\Lambda}}
 + \frac{a_2 + a_3\log(\log\frac{p}{\Lambda})}{\log^2\frac{p}{\Lambda}}
 + \mathcal{O}\,(\frac{1}{\log^3\frac{p}{\Lambda}})\,.
\end{equation}
Numerical coefficients $a_i$ may depend on the scheme used, but in a trivial
way: all the difference between $\{a_i\}$ and $\{\tilde a_i\}$ is
eliminated by a proper rescaling
$\Lambda\rightarrow c\,\Lambda$\,, with $c$ constant.
For instance, imposing $a_2=0$ renders all other coefficients
scheme-independent. In (\ref{expans}), $\Lambda$ is treated as
a formal expansion variable, the same for each scheme. However,
(\ref{Lambda}) and (\ref{psi1}) explain why $\Lambda$ is suited here:
by definition, it behaves like an (almost) scheme-independent quantity,
$\tilde\Lambda=\mu\exp(-\tilde\psi(\tilde g\,(g)))
=\mu\exp(-\psi(g)+\text{const})=c\,\Lambda$\,.

\vspace{.5cm}

\noindent
{\small {\bf Acknowledgements}. This work was supported by the RFBR
grant 00-01-00299.}

\end{document}